\documentstyle[preprint,eqsecnum,aps]{revtex}
\begin{document}
\draft
\preprint{Submitted to {\it Applied Physics Letters}}
\title{Scanning Tunneling Microscopy and Tunneling Luminescence of
       the Surface of GaN Films Grown by Vapor Phase Epitaxy}
\author{B. Garni, Jian Ma, N. Perkins, Jutong Liu, T.F. Kuech,
        and M.G. Lagally}
\address{University of Wisconsin -Madison, Madison, WI 53706}
\date{\today}
\maketitle

\begin{abstract}
We report scanning tunneling microscopy (STM) images of surfaces of
GaN films and the observation of luminescence from those
films induced by highly spatially localized injection of electrons or
holes using STM. This combination of scanning tunneling luminescence (STL)
with STM for GaN surfaces and the ability to observe both morphology and
luminescence in GaN is the first step to investigate
possible correlations between surface morphology and optical properties.
\end{abstract}

\narrowtext
\newpage
Gallium nitride (GaN) is of great interest because of its potential as a
material for the development of light-emitting devices in the blue and
ultraviolet spectral region.\cite{Davis,Pankove,Akasaki}
Extensive investigations have been done on GaN from different
perspectives, most of which have been focused on
the exploration of appropriate substrates and growth
conditions needed to yield materials
suitable for electroluminescent devices. Films of GaN have been
grown on various substrates by several techniques including
molecular beam epitaxy (MBE),\cite{Lacklison} and  different forms of
vapor phase epitaxy (VPE).\cite{Maruska,Qian}

GaN normally crystallizes with the hexagonal wurtzite structure
($\alpha$-GaN)\cite{Pankove,Elwell}, although it can also be synthesized
in its metastable cubic zinc-blende phase ($\beta$-GaN).
The techniques utilized in characterizing GaN include X-ray
diffraction, I-V characteristics, photoluminescence, reflection high-energy
electron diffraction (RHEED), and Auger electron spectroscopy (AES).
For both MBE and VPE growth, we still lack an understanding of
the surface morphology of GaN, especially the initial stage of nucleation and
its impact on subsequent film growth. Additionally,  little is known about
the intrinsic properties of GaN surfaces, restricting our ability to
control the surface composition and structure during processing.

Previous studies of GaN surfaces were limited to the employment of
RHEED and AES\cite{Hunt} to evaluate long-range order during
film growth, and to the study of the effect of sputtering/annealing on
N depletion. Recently, atomic force microscopy (AFM) observations
of GaN nucleation layers\cite{Kapolnek} and of the influence of open core
screw dislocations on the surface morphology of GaN films\cite{Qian}
have been reported.
However, there have as yet been no reports on the use of scanning
tunneling microscopy (STM) or scanning tunneling luminescence (STL)
measurements for GaN.

In this Letter, we report on STM and STL studies of
GaN films. Our results demonstrate the use of STM to
investigate the surface morphology of GaN. The observed
luminescence establishes the use of STL
to probe the local surface optical properties of GaN.

STL has recently been proven to be an effective tool in probing local
optoelectronic properties.\cite{Gimzewski,Abraham,Alvarado,%
Renaud,Ushioda,Samuelson,Sivel}
Our STM and STL experiments were performed in a custom-built STM/STL
UHV chamber\cite{Garni} with a base pressure of better
than $1 \times 10^{-10}$ torr.  STM/STL tips are made from 0.25 mm
diameter tungsten wire by means of electrochemical etching.
In STL, the tip is used as a highly localized source of electrons or holes to
induce luminescence in the material. The emitted photons are collected by
a lens mounted on the microscope base, positioned at 60$^\circ$ with respect
to the sample surface normal, with a solid angle of collection of
approximately 1 steradian. The collected light is directed to a lens
outside the chamber that focuses it onto a cooled GaAs cathode
photomultiplier tube (PMT). The spectral range for our collection/detection
apparatus is  approximately 320 nm - 900 nm,
and $\sim$0.8\% (at 360 nm) of the total emitted light from the surface can be
transmitted to the PMT. Photon counting continues while rastering the tip
in a constant-current tunneling mode, allowing both topological and position
dependent luminescence data to be acquired simultaneously, giving the
capability to correlate luminescence intensity with surface morphology.

GaN samples were grown on (0001) sapphire in a 60 mm diameter horizontal
hot-wall two-temperature-zone
halide vapor phase epitaxy (HVPE) reactor.\cite{Maruska}
In the first zone, typically operated around 850 C, the
reaction of MBE-grade 99.9999\% Ga metal with high
grade HCl gas in a N$_2$ carrier  gas results in GaCl and H$_2$  gas
reaction products. These products are transported downstream to the second
reaction zone, where NH$_3$ is introduced. The reaction of GaCl with
NH$_3$ produces GaN on the substrate surface, liberating HCl
and H$_2$ as reaction byproducts. This second zone is usually operated
at $\sim$1040 C. Films used for this
study were produced under a total flow of 10 standard liter per minute,
of which 880 standard cc per minute (sccm) was NH$_3$ and 29.4 sccm was
HCl. The rate of film growth is about 1.8 $\mu$m per minute.
Following  the growth, the sample is slowly withdrawn from the hot zone
under an NH$_3$ atmosphere until
the sample has  cooled. The reactor is then purged under N$_2$, and the
sample is removed for characterization. Epitaxial $\alpha$-GaN films grown
in this manner are naturally n-type, with a (0001)
orientation. The films have a small azimuthal mosaic spread of
$\sim$ $\pm$200 to 450 arcsec. The dislocation density is expected to be
high.\cite{Qian} The films have a band gap of 3.37 eV, as determined
by our separate photoluminescence
measurements. They exhibit a low reactivity to surface
contamination or chemical modification, and are stable enough to be
transported from the growth chamber to other chambers for
characterization without the need for extensive surface cleaning.
X-ray photoelectron spectroscopy studies performed after such transport
show monolayer amounts of C and O contamination.\cite{Ma}

Figure \ref{image_STM} shows  $1650 \times 1650$ \AA\  (a) and
$270 \times 270$ \AA\  (b) STM images of the surface of a nominally
$\sim$20 $\mu$m thick GaN(0001) film. In order to prevent inadvertent
modification of the as-grown surface, we performed no {\it in situ} cleaning
in the STM chamber, relying on the natural passivity of the
GaN surface to allow STM imaging.
The images in Fig. \ref{image_STM} were taken at a tip bias voltage of 4.8 V
with a tunneling current of 10 nA resulting in the injection of holes
into the near-surface region. A high bias voltage is needed for
imaging because GaN has a fairly wide band gap.  At low
magnification [Fig. \ref{image_STM}(a)], we observe an array of relatively
flat areas with seemingly random edges.  The high-magnification image
shown in Fig. \ref{image_STM}(b) reveals the roughness of the GaN surface
at the near-atomic scale. The bright flat areas correspond
to the topmost GaN layers; they have sizes between $\sim$50 and 120 \AA\ .
The height between two adjacent terraces varies between 10 and 20 \AA,
which corresponds to 4 - 8 atomic layers. We obtain similar images with
bias voltages of 4.8 V, 6.0 V, and 9.0 V, although resolution
is reduced at the higher bias voltages. At this stage, we have not yet
obtained  atomic resolution, most probably because of the submonolayer of
C and O contamination.\cite{Ma}

Luminescence measurements were made in two ways. In the cathodoluminescence
mode, the tip is retracted to $\sim$1 $\mu$m above the sample surface
and a voltage of $\sim$150 V
is applied between tip and sample, producing a beam of electrons
that easily overcomes the tunneling barrier, but nevertheless is
spatially very localized. In the tunneling mode, the tip is in the usual
position for topographic measurements, $\sim$10 \AA\ from the surface,
the bias voltage is of the order of several volts, and  electrons tunnel
either into or out of the solid.
The spatial extent at the point of injection is the same as in STM.
Fig. \ref{lumin_A-V} shows the luminescent intensity for a GaN film in
both cases. Fig. \ref{lumin_A-V}(a) shows the dependence of
the cathodoluminescent intensity on field emission current.
The tip is held at a constant distance. Small changes in the tip-sample
voltage (here 143 - 151 V) change the field emission current. A linear
dependence on beam current is observed, as expected.
We obtained similar current-luminescence profiles in different
regions of the sample. We have so far not measured the spectral
distribution of the cathodoluminescence or the tunneling luminescence,
but assume that most of the light comes in the band (364 nm) expected for GaN.

Figure \ref{lumin_A-V}(b) shows the dependence of the luminescent
intensity on the tip voltage in the tunneling mode with the sign of
the bias corresponding to hole injection. The measurement is made
at fixed lateral position of the tip, but with changing vertical position
so as to maintain a constant tunneling current as the voltage is varied.
The figure shows an average over 12
measurements; the error bars indicate the variation in the measurements.
A simple-minded explanation of the process of the tunneling luminescence
may be obtained in the following manner.\cite{Renaud}
The tunneling current will mimic the joint density of states (JDOS) of the
GaN valence band and metal tip valence band as electrons tunnel into the
metal tip. The holes in GaN will rapidly
thermalize to the top of the valence band. Luminescence occurs through
recombination of the holes with electrons at the bottom of the conduction
band. With those highly doped
n-type samples,  the Fermi level ($E_f$) will be very near the
conduction band edge, and we should observed a threshold of luminescence at
$\sim$3.37 eV (the band gap of GaN). Some luminescence actually begins to
appear above $\sim$2 eV, suggesting possible deep-level electron trap
states in the band gap, in agreement with our photoluminescence
measurements.\cite{Liu} Near the threshold, the luminescence intensity has a
$V^2$ dependence, consistent with STL observations
on Al$_x$Ga$_{1-x}$As.\cite{Renaud}.
The intensity beyond the threshold increases, but dips slightly near 6.0 eV.
Calculations of the density of states of GaN, as well
as photoemission measurements,\cite{Hunt} give a shape that has a local
minimum 3.5 eV below the top of the valence band. The
dip we observe in the luminescence corresponds qualitatively to this dip in the
valence band DOS. Beyond about 8.0 eV, the intensity shows a monotonic
increase, as we begin to reach the field emission region, as for
cathodoluminescence.

We have also made two-dimensional maps of the tunneling luminescence.
Because the intensity is weak, the images are still quite noisy.
Although the luminescence images show features that are reproducible from
one scan to the next, we see at this stage no obvious correlation
to the morphology observed by the STM. The resolution of STL relative
to STM is an open question. Because the mean free path of electrons and holes
before recombination may be large, the spatial resolution of STL at
first glance is expected to be considerably poorer than that of STM.
On the other hand, our calculations show that a high recombination site
density at the surface or at defects may greatly enhance the resolution
of the luminescence.\cite{Liu} At this stage we also do not know to what
extent the slight surface contamination affects the luminescence. We are
continuing to assess these questions.

In summary, we have succeeded in imaging GaN films with
STM and detecting its localized luminescence with STL for the first time.
The step morphology of the surface has been observed. The success of STL
experiments on GaN opens up the possibility of studying
correlations of surface morphology and defect structure with
local optical properties of GaN.

This work was supported by ONR, Physics and by NSF Grant No. DMR.91-21074,
the Naval Research Laboratory, and the ARPA URI on Visible Light Emitters.

\begin{figure}
\caption{Scanning tunneling microscopy (STM) images of the surface of
VPE grown GaN(0001) films. (a) 1650 $\times$ 1650 \AA\ .
(b) 270 $\times$ 270 \AA\ . The data were taken with the tip
biased at 4.8 volts with respect to the sample
with a constant tunneling current of 5 nA.}
\label{image_STM}
\end{figure}

\begin{figure}
\caption{Spatially localized luminescence intensity from GaN(0001).
(a). Cathodoluminescence using field emission from the STM tip. Dependence
of the luminescent intensity on the emission current. The tip is
1 $\mu$m from the sample and its potential is $\sim$ -150 V with respect to
the sample. (b). Tunneling mode with hole injection. Dependence of
luminescence intensity on the tunneling voltage at a constant tunneling
current.}
\label{lumin_A-V}
\end{figure}

\end{document}